\documentclass[preprint,12pt]{elsarticle}

\usepackage{graphicx}
\usepackage{amsmath}
\usepackage{amssymb}
\usepackage{amsthm}
\usepackage{booktabs}
\usepackage[update,prepend]{epstopdf}
\usepackage[slantedGreek]{mathptmx}
\usepackage{url}
\usepackage{bbm}
\usepackage{accents}
\usepackage{subfigure}

\newlength{\dhatheight}

\renewcommand{\div}{\mbox{\rm div}}

\newcommand{\xb}{\bar{x}}
\newcommand{\ab}{\overline{{\rm BR}}}
\newcommand{\br}{{\rm BR}}

\newtheorem{theorem}{Theorem}

\newtheorem*{theorem*}{Theorem}

\journal{NaN}

\begin{document}

\begin{frontmatter}

\title{ Learning by Fictitious Play in Large Populations}
\author{Misha Perepelitsa }

\date{\today}
\address{
misha@math.uh.edu\\
Department of Mathematics\\
University of Houston\\
4800 Calhoun Rd. \\
Houston, TX.}

\begin{abstract}
We consider learning by fictitious play  in a large population of agents engaged in single-play, two-person rounds of a symmetric game, and derive a  mean-filed type model  for the corresponding  stochastic process. Using this model, we describe qualitative properties of the learning process and discuss its asymptotic behavior. Of the special interest is the comparative characteristics of the fictitious play learning with and without a memory factor.  As a part of the analysis, we show that the model leads to the  continuous, best-response dynamics equation of Gilboa and Matsui (1991), when all agents have similar empirical probabilities.

\end{abstract}

\begin{keyword}
Fictitious play \sep best response dynamics \sep learning in large populations 



\end{keyword}

\end{frontmatter}

\begin{section}{Introduction}

Learning theories concern with the rules by which  players can  discover  optimal strategies in repeated plays of games, typically, when the players act in self-interest,  in the absence of the complete information about the game, and having limited ability to communicate with other players.


In the learning by fictitious play (FP) one assumes that players keep the statistics of their opponent's actions
over the whole history of the process and compute  empirical probabilities for actions played, as if playing against  stationary environment. Given the game payoffs,  agents' actions are the best responses to  their assessment of opponents. A number of sufficient conditions for convergence of the empirical probabilities to Nash equilibria was established by Robinson (1951), Miyasawa (1961), Nachbar (1990), Krishna \& Sjostrom (1995).

Fictitious play learning, however, need not to converge, as shown by an example of  Shapley (1964), in which empirical probabilities follow a cycle. This type of behavior was discussed in greater details by Gilboa \& Matsui (1991), Gaunersdorfer \& Hofbauer (1995),  Monderer et al.  (1997), and  Benaim, Hofbauer \& Hopkins (2009).

The convergence of empirical probabilities, even if it does take place, does not tell the whole story of learning. Equally important is to know what actions are selected by the payers in the course of learning. An example of Fudenberg \& Kreps (1993), for the game in Table \ref{table:test}, shows that the process can go through the correlated play of $(L,L),\,(R,R),\,(L,L),\,(R,R)...,$ with players realizing zero payoffs, rather than positive payoff of the Nash equilibrium. A variant of fictitious play, called  stochastic fictitious play, was introduced by Fudenberg \& Kreps (1993), following the idea of  Harsanyi's (1973), to provide a reasonable learning model in which players choose mixed strategies as their best responses.
The convergence of stochastic FP in 2x2 games in various situations was established by
Fudenberg \& Kreps (1993), Benaim \& Hirsch (1996) and Kanoivski \& Young (1995).

There are different scenarios for learning to evolve in a population of agents depending on available information and how plays are arranged between agents. One of the scenarios, which we adopt in this paper, is to consider sequential plays   between pairs of randomly selected agents, and keep  the outcomes of the games  private. To describe learning in such processes, Gilboa \& Matsui (1991) proposed the continuous, ODE model for the change of the distribution of players on a set of pure strategies. A similar equation was used by Fudenberg \& Levine (1998) to describe the changes in population average subjective probabilities.

The equation holds under conditions that the number of players is large, and only small number of players are adjusting their play to the best response of the population average during short time periods. It is also implicitly assumed that all players in the population have nearly similar vector of empirical probabilities at all times and the best response function is evaluated at that vector.

The equation, known as the best response dynamics (BRD) equation, became popular model
for studying FP learning in large populations, and its asymptotic properties were discussed by   Hofbauer \& Sigmund (1998),  Gaunersdorfer \& Hofbauer (1995), Benaim, Hofbauer \& Hopkins (2009).

In this paper we would like to obtain a refinement of the BRD equation by considering 
the changes in the probability density function for the distribution of agents in the space of 
empirical priors, rather than postulating equations for its moments. 

The mean-field model, that we obtain, contains significantly more information about the state of the learning. It shows  the whole spectrum of empirical priors in the population, and indicates how learning affects agents with different subjective probabilities. 

We follow the approach, based on Fokker-Planck equation,  is classical in many-agents systems in physics, biology, and social sciences.  It applies to systems in the state of ``chaos" when states of the learning of two randomly selected agents are independent (or nearly independent). This condition is reasonable in large populations where the same pair of agents is rarely selected for the play. At the same time, it excludes any type of correlation (coordination) between players.

Two models of fictitious play are considered in this paper: the classical FP and FP with  a memory factor. The latter model places higher weights to  more recent observations, compared to equally weighted classical model. Memory factor models have been used in the context of  reinforced learning by Harley (1981), Erev \& Roth (1998), Roth \& Erev (1995)  and in fictitious play learning by Benaim, Hofbauer \& Hopkins (2009).

The paper organized as follows. In section \ref{model} we describe the model and the PDE  equations that serve as an approximation of the learning process. The derivation of the equations from many-particle stochastic process is presented in Appendix. Section \ref{test} compares the dynamics of the stochastic learning and the corresponding dynamics from the deterministic PDE model, on the example of the 2x2 miscoordination game from Table \ref{table:test}. In section \ref{RBD} we explain how BRD equation is obtained from the PDE model, and in sections \ref{memory} and \ref{memory} consider the learning dynamics obtained from the PDE model for learning in games with a single Nash equilibrium. In section \ref{2x2} we return to the example of the  miscoordination game of Fudenberg \& Kreps (1993), and using the PDE model as the predictor for the evolution of learning, we show that both, mean subjective probabilities and mean best response probabilities, converge to the Nash equilibrium. Thus, in large populations, in roughly half of the plays each agent gets a positive payoff.

\begin{table}
\centering
\begin{tabular}{@{}lcc@{}}
\toprule
    &  L         & R            \\
L  & (0,\, 0)   & (1,\,1)    \\
R  &  (1,\, 1) &  (0,\, 0)  \\
\bottomrule
\end{tabular}

\vspace{15pt}

\caption{Persistent miscoordination game from Fudenberg \& Kreps (1993).  \label{table:test}}

\end{table}

\end{section}

\begin{section}{The Model}
\label{model}
We consider a series of plays of a symmetric 2-player game between randomly selected agents in a large population. Denote the set of pure strategies in the game by $\{s_i\},$ $ i=1..n.$  Each agent $k$   maintains a record $(S^t_{1,k},..,S^t_{n,k})$ of 
times her opponents played  $(s_1,..,s_n)$ up to the epoch $t.$ 

Agent $k$ is using this vector as a prior for the probability
\[
\frac{S^t_{i,k}}{\sum_j S^t_{j,k}}
\]
 for action $j$ to be played 
next time by her opponent, and selects her action as a best response. In the classical fictitious play the best response is 
a multi-valued function. For definiteness we assume that there is a rule by which agent $k$ decided between equivalent actions. This will not enter in the equations  for the averaged dynamics, as long as the situation is non-generic: the set of strategies for which 
the best response is multi-valued is of measure zero. The latter condition is assumed in the paper.

After the play, agent $k$ updates her empirical priors by the rule
\begin{equation} 
\label{rule}
S^{t+1}_{i,k}{}={}(1-\mu h) S^t_{i,k} + I_{i,k},\quad i=1..n,
\end{equation}
where $I_{i,k} = h$ if the opponent played $s_i,$ and zero otherwise. The  learning increment $h$ can be any positive number, without altering large time learning process.  A positive parameter $(\mu h)\in[0,1]$ is the memory factor. Leaning starts with agents having initial priors $(S^0_{1,k},..,S^0_{n,k}).$ Two agents are selected at random for a play of the game each epoch $t.$ 
Only  the agents who played the game update their priors, and the outcome is not revealed to others.  The stochastic process defined in this way is a discrete-time Markov process on the $nN$ dimensional state space of priors.

\subsection{Priors state space}

Our main interest is in the probability density function (PDF) $f(x,t)$ of agents in the priors-space $x=(x_1,..,x_n),$ $x_i\geq0,$ where $x_i$ is the number of times (actual count, not the proportion) action $i$ was played by an agent's opponent.  In this space the straight lines through the origin represent the sets of the opponent's constant  probabilities to play $(s_1,..,s_n),$ that is, the vector of empirical probabilities is $x/\sum_j x_j.$ For any subset $\Omega$ in the priors-space, $\int_\Omega f(x,t)\,dx$ represents the proportion of agents with their priors in the set $\Omega$ at time $t.$ By $\br=\br\left(\frac{x}{\sum_jx_j}\right)$ we denote the best response (vector) function. We consider the generic case when $\br(x)$ takes on $n$  values $\{e_k\}_{k=1}^n$ of unit, basis vectors,
away from a set of measure zero. In large populations $f(x,t)$ is well approximated by a bounded function (not a distribution) and the value of the population mean best response
\begin{equation}
\label{rb}
\ab(t){}={}\int \br\left(\frac{x}{\sum_jx_j}\right)f(x,t)\,dx
\end{equation}
does not depend on the values of $\br$ on the exceptional set.

Our main interest is in describing the changes in $f(x,t)$  in the process of learning and analyzing its asymptotic behavior. Here, the main numerical characteristics  are the vector of mean empirical probabilities
\[
\Lambda(t){}={}\int \frac{x}{\sum_j x_j}f(x,t)\,dx,
\]
and  the mean best response vector $\ab(t).$

The following equation is found (see Appendix) to be the leading order approximation of the stochastic learning process, when  the number of players is large and the learning increment $h$ is small.
 \begin{equation}
\label{eq:main}
\frac{\partial f}{\partial t}{}+{}\div (u(x,t)f){}={}0,
\end{equation}
with  velocity $u$ given by the formula 
\begin{equation}
\label{velocity}
u(x,t){}={}\ab(t) - \mu x.
\end{equation}

We will refer to equations \eqref{rb}, \eqref{eq:main}, and  \eqref{velocity} as the PDE model of the fictitious play learning. To complete the description of the model we will prescribe zero boundary conditions for function $f.$ Because the domain $x_i\geq0,$ $i=1..n,$ is invariant under the flow of \eqref{velocity}, the problem is correctly posed (not over-determined). Notice also that the problem is non-linear: the ``conservation of mass'' equation \eqref{eq:main} carries the density  $f$ with velocity $u(x,t)$ which, in its turn, depends on  function $f.$

\subsection{Numerical test}
\label{test}
In this section we compare the solution of the PDE model for a 2x2 game with a mixed Nash equilibrium, given in Table \ref{table:test}, with the direct simulation of the learning process.

We take the initial data $f_0$ is the density of uniform distribution of initial priors in the box
$B=[0,1]\times[3,4].$ The number of agents is  $N=1000$ and the learning increment is $h=0.001.$ The memory factor is set to zero. The game has a single mixed Nash equilibrium $(1/2,1/2).$ We consider the state of the learning at time $t=20$ when the learning settles near the equilibrium. By formula \eqref{time} this corresponds to $20N/(2h)=1000/h$ iterations of the game.

Recall that at all times the solution of the PDE is an uniform distribution on a box of side-length $1,$ and only the coordinates of the center of the box, $x_c,$ need to be computed. We use the explicit Euler method for the ODE
\[
\frac{d x_c}{dt} = \ab(t){}={}\int_{B(x_c)}\br\left(\frac{x}{\sum_j x_j}\right)\,dx,
\]
and find that at  $t=20,$  $x_c=(12.001, 12.001),$ see Figure \ref{fig:test2}. We compare it with the data points that are the priors of 1000 agents after $1000/h$ random plays. At time $t=0,$ the priors of agents had been selected from an uniform distribution on the box $B.$  The figure shows good agreement of the PDE model with the actual learning process.

\begin{figure}[t]
\centering
\includegraphics[width=8cm]{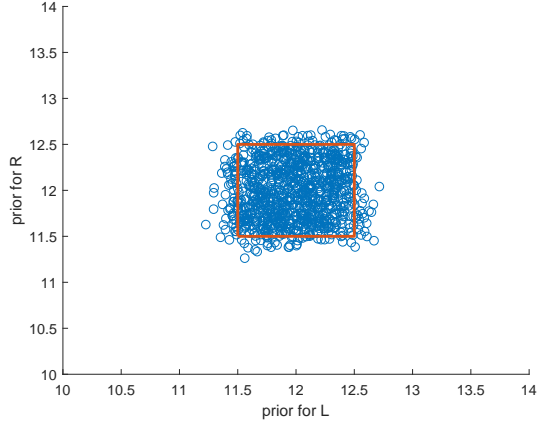}
\caption{Distribution of priors. Uniform distribution on the unit box (shown in red) is predicted by the PDE model. The center of the box, point (12,001,12,001), corresponds to the Nash equilibrium in the game. Data points are priors from learning in the population of N=1000 agents in a single run of the model for 40000 iterations. At time $t=0$ distribution of priors is uniform on a box $[0,1]\times[3,4]$ (not shown).
 \label{fig:test2}}
\end{figure}

\subsection{Relation to the Best Response Dynamics (BRD)  equation }
\label{RBD}
Using equation \eqref{eq:main}, one can compute the equation for the mean empirical frequencies vector $\Lambda(t):$
\begin{equation}
\label{Lambda}
\frac{d\Lambda_i}{dt}{}={}\int \frac{1}{\sum_j x_j}\left(\ab_i(t) - \frac{x_i}{\sum_j x_j}\right)f(x,t)\,dx,\quad i=1..n,
\end{equation}
since $\sum_j \ab_j(t){}={}1.$ Note, that the memory factor $\mu$ does not explicitly enter the equation for $\Lambda.$ It does, however, contributes to the dynamics of  the distribution $f.$

If one postulates that all agents have the same, or approximately the same, priors 
\begin{equation}
\label{H:delta}
x(t)=(x_1(t),..,x_n(t)),
\end{equation}
then $f(x,t)$ is concentrated near $x(t)$  and the above equation reduces  to a variant of  the best response  dynamics equation:
\begin{equation}
\label{BRD}
\frac{d\Lambda_i}{dt}{}={}\frac{1}{\sum_j x_j(t)}\left( \widetilde{\br}_i(\Lambda) -\Lambda_i  \right),\quad i=1..n.
\end{equation}
In this equation $\widetilde{\br}$ is ``a regularization'' of the best response function $\br$ over the support of function $f(x,t).$ If the latter
converges to a delta mass, $\widetilde{\br}$ converges to a value of $\br(\Lambda).$ 
Notice, also, the  positive factor on the right-hand side of the equation. For a learning processes in which priors become large,  the learning rate slows down. The learning factor corresponds to the factor $t^{-1}$ in  BRD equation in Fudenberg \& Levine (1998). 

Hypothesis \eqref{H:delta} can be replaced with a weaker one, by requiring that the empirical probabilities $x(t)/(\sum_j x_j)$ in the support of $f(x,t)$ are nearly constant. 
The extent to which this hypothesis or  \eqref{H:delta} are consistent with the dynamics of \eqref{eq:main} is limited only to the case when $u(x,t)$ has a single, asymptotically stable fixed point, or the support of $f$ is bounded but it is carried by the velocity to large values of $|x|,$ so that the empirical probabilities inside the support are nearly constant. The former condition holds for the model with a memory factor $\mu>0,$ and latter for the  model with $\mu=0.$

\subsection{Fictitious play with memory factor}
\label{memory}
In this and the following sections we will assume that the problem \eqref{rb}--\eqref{velocity} has a unique, regular solution with   a generic initial data $f(x,0)=f_0(x),$ the fact that can be proved by standard PDE techniques. To start with the qualitative analysis of the model with $\mu>0,$  notice that the velocity is a simple linear function of $x,$  and for all $i,$ $\partial_{x_i}u(x,t){}={}-\mu<0.$ A flow of this type compresses the support of $f$ (the set where $f$ is positive) toward a point, the position of which, in general, changes. Notice also that each component $u_i(x,t)$ becomes negative when $x_i$ is sufficiently large. This means that the learning takes place in the bounded region of the priors space. After some transient time, all mass of $f$ is concentrated near a point and the dynamics is approximated by equation \eqref{BRD}, where the factor $1/\sum_j x_j$ is larger than some fixed positive number.  The asymptotic behavior is determined by the BRD equation \eqref{BRD}. In the presence of a dominant strategy, the solution of the PDE model will converge to a delta mass concentrated at  a boundary point, located on the boundary set $\{x\,:\, x_i{}={}0\},$ corresponding to the dominant strategy $s_i.$

For symmetric games with the payoff matrix 
$
A=\{a_{ij}\}_{i,j=1}^n,
$
 such that the quadratic form $x\cdot Ax$ is strictly concave on a strategy simplex 
 \[
 \Delta {}={}\left\{x\,:\, x_i\geq 0,\,i=1..n,\,\sum_i x_i=1\right\},
 \]
  Hofbauer (2009)  showed that solutions of BRD (or smoothed BRD) equation converge to a unique Nash equilibrium. With marginal modifications one can show that solutions of \eqref{BRD} converge to a unique Nash equilibrium as well. Combined with the fact  the support of $f$ converges to a point, we conclude that for all agents will have the same vector of the equilibrium empirical probabilities.

We stress again that above arguments are based on the leading order approximation  of the stochastic learning process. The next order is the drift-diffusion equation  \eqref{FP:multi} from Appendix. In that model, the outcome of the learning will be a stationary distribution of small deviation of order $h$ around the Nash equilibrium.

\subsection{Fictitious play with zero memory factor $\mu=0$}
\label{nomemory}
In this case the velocity $u(x,t)$ is uni-directional: $u(x,t)=\ab(t).$ The initial profile of the probability distribution $f_0(x)$ is simply carried by the velocity  and its shape doesn't change. All components of velocity $\ab(t)$ are non-negative and $|\ab(t)|\geq n^{-1},$ as can be seen from  formula \eqref{rb}. With such velocity $f$ is moved away from the origin at non-vanishing speed.

 Suppose that at $t=0,$ $f_0$ is an arbitrary distribution on a box $B$ of side-length $a.$  After time $t,$ $f(x,t)$ is given by the same distribution on a box that is located, approximately, $t/n$ units away from the origin. For large $t$ the vector of empirical frequencies $x/\sum_jx_j$ is approximately constant in the ball (with deviations of order $at^{-1}$), effectively rendering it as a single point. The dynamics can now be computed from the equation \eqref{BRD}, in which the learning factor $(\sum_j x_j)^{-1}$ is of the order $t^{-1}.$  Under the strict concavity of on the expected  payoff $x\cdot Ax,$ the vector of average  
empirical frequencies  will converge to a unique Nash equilibrium.


\subsection{Equation for the mean best response: a 2x2 game}
\label{2x2}
Equation \eqref{BRD} and the BRD equation of Gilboa \& Matsui (1991) are the equations for the population averaged empirical probabilities.  Another important characteristic of a learning process is the average
strategy played at time $t.$ In this section we show on a simple example that PDE model
\eqref{eq:main} can be used to derive the equation for $\ab(t).$
Consider the game in Table \ref{table:test} and FP learning  model with zero memory factor $\mu=0.$ We will take initial distribution of priors to be uniform on some box. 
As was mentioned earlier, the learning dynamics (the leading order) transports the initial distribution $f_0$ with uni-directional velocity $\ab(t).$ If at time $t$ the support of $f(x,t)$ is completely contained in the wedge $x_1>x_2$ (or $x_2>x_1$), the velocity is constant $(0,1)^t$ (or $(1,0)^t$). This type of velocity moves $f_0$
towards the line $x_1=x_2.$ When the support of $f(x,t)$ intersects that line, we denote by $l(t)$ the length of segment of intersection, and by multiplying \eqref{eq:main} by function $\br$ and integrating over the whole domain we obtain the following first order system of ODEs for $\ab(t):$
\[
\frac{d\ab}{dt}{}={}l(t)\left[\begin{array}{rr}
                                         - 1& 1\\
                                         		1& -1
                                         		\end{array}	\right]\ab.
                                         		\]
 Assuming that after some transient time, $l(t)>l_0>0,$ by changing to the new time variable (still labeled $t$), the system is reduced to the constant coefficient case. We find that there are eigenvalues: $\lambda = 0, -2$ and the corresponding eigenvectors: $(1/2,1/2)^t$ and $(-1/2,1/2)^t.$ This implies that asymptotically,
 \[
 \ab(t){}={}    \left[\begin{array}{r}
                                         1/2\\
                                         	1/2
                                         		\end{array}	\right] + O(e^{-2t}).
                                         		\]
Mean population best reply strategy approaches the mixed Nash equilibrium.
With this information at hand, we can use equation \eqref{Lambda}, and by estimating the learning factor $\sum_jx_j$ by $t,$ obtain the following equation for the mean empirical probabilities $\Lambda:$
\[
\frac{d\Lambda}{dt}{}\approx{}\frac{1}{t}\left( \ab(t) -\Lambda\right).
\]
By solving it, we conclude that $\lim\Lambda(t)= (1/2,1/2)^t.$

The arguments leading to the convergence of $\Lambda(t)$ and $\ab(t)$ to the Nash equilibrium can be repeated for a generic $2x2$ symmetric game. Let the payoffs for actions $(L,L),\,(R,L),\,(L,R),\,(R,R)$ be $(a,a),\, (c,d),\,(d,c),\,(b,b),$ respectively, with $a<c,$ $b<d.$ The following theorem holds.
\begin{theorem}
 Suppose that the initial distribution of agents in the priors space is described by a smooth function $f_0$ with compact support in the open quadrant $x_1>0,\,x_2>0.$ Let $f(x,t)$ be the unique solution of the problem \eqref{rb}--\eqref{velocity}. Then, the mean empirical probabilities, $\Lambda(t)$,  and the mean best response vector, $\ab(t),$ converge as $t\to\infty$ to the unique, mixed Nash equilibrium. 
\end{theorem}

\end{section}





\begin{section}{Appendix: Fokker-Planck equation}

Consider a group of $N$ individuals acting according to FP learning rule described in section \ref{model} in a symmetric game. For the simplicity of the presentation, we restrict ourselves only to the case of two pure strategies $\{s_1,s_2\}.$ The model with $n$ strategies is written down at the end of the Appendix. Denote  $X^t_i{}={}(S^t_{1,i},S^t_{2,i}),$ the vector of counts of opponent's plays of $s_1$ and $s_2$ for  agent $i,$ up to epoch $t.$ By $X^t=(X^t_1,..,X^t_N)$ we denote the vector of counts for all agents.  By $w_h(\xb,t),$ where $x\in[0,1]^{2N},$ we denote PDF for distribution of $X^t.$ We will write $\xb=(x_1,..,x_N),$ where each $x_i{}={}(s_{1,i},s_{2,i}).$ The best response of agent $i$ will be denoted as $\lambda_i{}={}\br_i \left(x_i/(s_{1,i}+s_{2,i})\right).$ Suppose that agents $i$ and $j$ are selected for the interaction. There will be only one game played during the period from $t$ to $t+\delta.$ 

We consider the learning rule \eqref{rule} in which the priors are incremented by $h>0,$ that is, $I_{i,k}$ is either $h$ or $0.$ Since the best response function depends on the empirical probabilities rather than on priors, the magnitude of the increment is irrelevant, apart from the fact that for smaller increments the initial data influence the dynamics for longer periods of time. The memory factor in \eqref{rule} is set to $\mu h,$ with $0\leq \mu h\leq 1.$  The games are arranged to be played at time periods of length $\delta$ determined  as
\begin{equation}
\label{time}
\delta{}={}\frac{2h}{N}.
\end{equation}
 If $T$ is a time scale of learning process (in arbitrary units), then $T/\delta$ is the number of rounds of the games needed to be played to observe changes at this time scale. All games are played between two agents, thus, on average an
 agent plays $T/(\delta N)$ times during an interval length $T.$
 
Conditioned on the event $X^t=\xb,$ the agents priors for the next period are set according to formulas
\[
X^{t+\delta}_i{}={}\left\{
\begin{array}{ll}
((1-\mu h)s_{1,i} + h,\, (1-\mu h)s_{2,i}) & \rm Prob=\lambda_i\lambda_j\\
((1-\mu h)s_{1,i},\,(1-\mu h)s_{2,i}+ h) & \rm Prob=\lambda_i(1-\lambda_j)\\
((1-\mu h)s_{1,i} + h,\,(1-\mu h)s_{2,i}) & \rm Prob=(1-\lambda_i)\lambda_j\\
((1-\mu h)s_{1,i},\,(1-\mu h)s_{2,i} +h) & \rm Prob=(1-\lambda_i)(1-\lambda_j)
\end{array}
\right.
\]
and symmetrically for $X^{t+\delta}_j.$
For all other agents, $X^{t+\delta}_k{}={}X^t_k$ for $k\not=i,j.$ The definition of $X^t$ makes it a discrete-time Markov process. We proceed by writing down the integral form of the Chapman-Kolmogorov equations and approximate its solution by a solution of the Fokker-Planck equation (forward Kolmogorov's equation), for small values of $\delta, h$ and large $N.$ This is a classic approach to stochastic processes, the details of which  can be found in Feller's (1957) monograph. 

The change of $w_h(\xb,t)$ from $t$ to $t+\delta,$ can be described in the following way. 
\begin{multline}
\int \phi(\xb)w_h(\xb,t+\delta)\,d\xb{}={}\mathbb{E}[\phi(X^{t+h})]\\
{}={}\sum_{i\not=j}(N(N-1))^{-1}\int\left(
\lambda_i\lambda_j\phi(\xb)\Big|_{x_i{}={}((1-\mu h)s_{1,i} +h,\, (1-\mu h)s_{2,i})\atop 
\,\,\, x_j=((1-\mu h)s_{1,j} +h,\, (1-\mu h)s_{2,j})}\right.\\
\left.
+ \lambda_i(1-\lambda_j)\phi(\xb)\Big|_{x_i{}={}((1-\mu h)s_{1,i},\,(1-\mu h)s_{2,i}+h)\atop
\,\,\, x_j{}={}((1-\mu h)s_{1,j} +h,\,(1-\mu h)s_{2,j})}\right. \\
\left. 
+(1-\lambda_i)\lambda_j\phi(\xb)\Big|_{x_i=((1-\mu h)s_{1,i} +h,\,(1-\mu h)s_{2,i}) \atop  \, \,\, x_j=((1-\mu h)s_{1,j},\,(1-\mu h)s_{2,j}+h)} \right. \\
\left. 
+(1-\lambda_i)(1-\lambda_j)\phi(\xb)\Big|_{x_i=((1-\mu h)s_{1,i},\,(1-\mu h)s_{2,i} +h) \atop \,\,\, x_j=((1-\mu h)s_{1,j},\,(1-\mu h)s_{2,j} +h)}\right)w_h(\xb,t)\,d\xb.
\end{multline}
This equation can be written in slightly different way:
\begin{multline}
\label{eq:int1}
\int \phi(\xb)w_h(\xb,t+\delta)\,d\xb
{}={}\int \phi(\xb)w_h(x,t)\,d\xb \\
{}+{}\sum_{i\not=j}(N(N-1))^{-1}\int\left(
\lambda_i\lambda_j[\phi(\xb)\Big|_{x_i{}={}((1-\mu h)s_{1,i} + h,\, (1-\mu h)s_{2,i})\atop 
\,\,\, x_j=((1-\mu h)s_{1,j} + h,\, (1-\mu h)s_{2,j})}-\phi(\xb)]\right.\\
\left.
+ \lambda_i(1-\lambda_j)[\phi(\xb)\Big|_{x_i{}={}((1-\mu h)s_{1,i},\,(1-\mu h)s_{2,i}+ h)\atop
\,\,\, x_j{}={}((1-\mu h)s_{1,j} +h,\,(1-\mu h)s_{2,j})}-\phi(\xb)]\right. \\
\left. 
+(1-\lambda_i)\lambda_j[\phi(\xb)\Big|_{x_i=((1-\mu h)s_{1,i} +h,\,(1-\mu h)s_{2,i}) \atop  \, \,\, x_j=((1-\mu h)s_{1,j},\,(1-\mu h)s_{2,j}+ h)}-\phi(\xb)] \right. \\
\left. 
+(1-\lambda_i)(1-\lambda_j)[\phi(\xb)\Big |_{x_i=((1-\mu h)s_{1,i},\,(1-\mu h)s_{2,i} +h) \atop \,\,\, x_j=((1-\mu h)s_{1,j},\,(1-\mu h)s_{2,j} +h)}-\phi(\xb)]\right)w_h(\xb,t)\,d\xb.
\end{multline}

\vskip 10pt

Denote the PDF of the distribution by 
\[
f_h(x,t){}={}\sum_k N^{-1}\int w_h(\xb)\big|_{x_k=x}\,d\xb_k,\quad x\in\mathbb{R}^2,
\]
where $\xb_k$ is a $2N-2$ dimensional vector of all coordinates, excluding $x_k.$
In statistical physics this function is also called one-particle distribution. In the formulas to follow we need to use two-particle distribution function 
\[
g_h(x,y,t){}={}\sum_{i\not=j} (N(N-1))^{-1}\int w_h(\xb)\big|_{x_i=x,\, \\x_j=y}\,d\xb_{ij},
\]
where $\xb_{ij}$ is the $2N-4$ dimensional vector of all coordinated excluding $x_i$ and $x_j.$ Function $g_h$ is symmetric in $(x,y)$ and is related to $f_h$ by the formulas
\[
f_h(x,t){}={}\int g_h(x,y,t)\,dx{}={}\int g_h(x,y,t)\,dy.
\]
The moments of function $f_h$ and $g_h$ are computed from the  moment of $w_h:$
\[
\int \psi(x)f_h(x,t)\,dx{}={}\sum_k N^{-1}\int \psi(x_k)w_h(\xb)\,d\xb,
\]
and 
\[
\int \omega(x,y)g_h(x,y,t)\,dxdy{}={}\sum_{i\not=j} (N(N-1))^{-1}\int \omega(x_i,x_j)w_h(\xb)\,d\xb.
\]
This follows from the definition of these functions.

Now we use \eqref{eq:int1} to obtain an integral equation of the change of function $f_h.$ For that select $\phi(\xb){}={}\psi(x_k),$ sum over $k$ and take average. We get
\begin{multline}
\int \psi(x)f_h(x,t+\delta)\,dx{}={}\int \psi(x)f_h(x,t)\,dx\\
{}+{}N^{-1}\sum_{i\not=j}(N(N-1))^{-1}\int\left(
\lambda_i\lambda_j[\psi((1-\mu h)s_{1,i}+h,\, (1-\mu h)s_{2,i}) - \psi(x_i)  \right.
\\ \left.
+\psi((1-\mu h)s_{1,j} + h,\, (1-\mu h)s_{2,j})-\psi(x_j)]\right. \\
\left. \right.\\ \left.
+\lambda_i(1-\lambda_j)[\psi((1-\mu h)s_{1,i},\,(1-\mu h)s_{2,i}+h)-\psi(x_i)\right. \\
\left. + \psi((1-\mu h)s_{1,j} +h,\,(1-\mu h)s_{2,j}) - \psi(x_j)] \right. \\
\left. \right. \\
\left. +(1-\lambda_i)\lambda_j[ \psi((1-\mu h)s_{1,i} +h,\,(1-\mu h)s_{2,i})-\psi(x_i) \right. \\
\left. + \psi((1-\mu h)s_{1,j},\,(1-\mu h)s_{2,j}+h) - \psi(x_j)] \right. \\
\left. \right. \\ \left. +(1-\lambda_i)(1-\lambda_j)[\psi((1-\mu h)s_{1,i},\,(1-\mu h)s_{2,i} +h)-\psi(x_i)   \right. \\
\left. + \psi((1-\mu h)s_{1,j},\,(1-\mu h)s_{2,j} +h) - \psi(x_j)] \right)w_h(\xb,t)\,d\xb.
\end{multline}
The right-hand side can be conveniently expressed in terms of the two-particle function $g_h:$

\begin{multline}
\label{int_f}
\int \psi(x)f_h(x,t+\delta)\,dx{}={}\int \psi(x)f_h(x,t)\,dx\\
{}+{}N^{-1}\int\left(
\lambda(x)\lambda(y)[\psi((1-\mu h)s^x_1 + h,\, (1-\mu h)s^x_2) - \psi(x)  \right.
\\ \left.
+\psi((1-\mu h)s^y_1 + h,\, (1-\mu h)s^y_2)-\psi(y)]\right. \\
\left. \right.\\ \left.
+\lambda(x)(1-\lambda(y))[\psi((1-\mu h)s^x_1,\,(1-\mu h)s^x_2+h)-\psi(x)\right. \\
\left. + \psi((1-\mu h)s^y_1 + h,\,(1-\mu h)s^y_2) - \psi(y)] \right. \\
\left. \right. \\
\left. +(1-\lambda(x))\lambda(y)[ \psi((1-\mu h)s^x_1 +h,\,(1-\mu h)s^x_2)-\psi(x) \right. \\
\left. + \psi((1-\mu h)s^y_1,\,(1-\mu h)s^y_2+h) - \psi(y)] \right. \\
\left. \right. \\ \left. +(1-\lambda(x))(1-\lambda(y))[\psi((1-\mu h)s^x_1,\,(1-\mu h)s^x_2 +h)-\psi(x)   \right. \\
\left. + \psi((1-\mu h)s^y_1,\,(1-\mu h)s^y_2 +h) - \psi(y)]
\right)g_h(x,y,t)\,dxdy.
\end{multline}
where $x=(s^x_1,s^x_2),$ $y=(s^y_1,s^y_2),$ and $\lambda(x){}={}s^x_1/(s^x_1+s^x_2),$ and similar for $\lambda(y).$
In the processes with large number of agents and random binary interactions, two-particle distribution function can be factored into two independent distributions:
\[
g_h(x,y,t){}={}f_h(x,t)f_h(y,t).
\]
With this relation, \eqref{int_f}, becomes a family of non-linear integral relations for the next time step distribution $f_h(x,t+\delta).$ 
Selecting time step proportional to $h$ so that $\frac{2h}{\delta N}\sim 1,$ and taking Taylor expansions up to the second order for the increment of the test function $\psi,$ we obtain the following Fokker-Planck equation
\begin{equation}
\label{FP}
\frac{\partial f}{\partial t}{}+{}\div (u(x,t)f)-h\sum_{i,j=1,2}\partial^2_{x_ix_j}\left(d_{ij}f\right){}={}0,
\end{equation}
where $x=(x_1,x_2),\,x_1,x_2>0$ and the drift velocity is given by the formula
$
u(x,t){}={}\ab(t) - \mu x,
$
where 
\begin{equation*}
\label{m}
\ab(t){}={}\int\br\left(\frac{x}{x_1+x_2}\right)f(x,t)\,dx.
\end{equation*}
$d_{ij}$ are elements of a symmetric, positive definite diffusion matrix $D$   computed by the formula:
\[
D{}={} \ab_1(t)(\mu x_1-1,\mu x_2)\otimes(\mu x_1-1,\mu x_2){}+{}\ab_2(t)(\mu x_1,\mu x_2-1)\otimes(\mu x_1,\mu x_2-1).
\]

Consider now the learning  from playing a symmetric game with $n$ pure strategies.  Denote the priors vector $x=(x_1,..,x_n).$ Then, the Fokker-Planck equation approximating the stochastic learning is 
\begin{equation}
\label{FP:multi}
\frac{\partial f}{\partial t}{}+{}\div (u(x,t)f)-h\sum_{i,j=1..n}\partial^2_{x_ix_j}\left(d_{ij}f\right){}={}0,
\end{equation}
$
u(x,t){}={}\ab(t) - \mu x,
$
where 
\begin{equation*}
\label{m:1}
\ab(t){}={}\int\br\left(\frac{x}{\sum_jx_j}\right)f(x,t)\,dx.
\end{equation*}
$d_{ij}$ are elements of a symmetric, positive definite diffusion matrix $D$   computed by the formula:
\[
D{}={}\sum_i \ab_i(t) (\mu x-e_i)\otimes(\mu x-e_i).
\]


%


\end{section}

\end{document}